# Nucleation and growth of germanium islands during layer exchange metal-induced crystallization

Shu Hu[1] and Paul C. McIntyre[1,2]‡

[1]Department of Materials Science and Engineering, Stanford University, Stanford, CA 94305
[2]Geballe Laboratory for Advanced Materials, Stanford University

E-mail:
pcm1@stanford.edu (Paul C. McIntyre)
‡ Author to whom any correspondence should be addressed.

**Abstract**

Al-induced layer exchange crystallization of amorphous Ge thin films has been demonstrated recently, and provides a suitable system to characterize, model and control Ge crystal growth on non-crystalline substrates. Direct observation of Ge transfer to the surface of Al through an interposed $GeO_x$ interfacial layer allows independent measurement of the density and average area of crystalline Ge islands formed on the surface. Based on these experimental observations, the Johnson-Mehl-Avrami-Kologoromov phase transformation theory is extended to model nanoscale nucleation and growth of Ge islands in two dimensions. The Ge island growth mechanism switches from atomic-attachment-limited to surface diffusion-limited kinetics with increasing time. The transition point between these regimes depends on the Ge nucleation site density and the annealing temperature. Finally, we show that local bias-voltage stressing of the interfacial layer controls the areal density of nucleated Ge islands on the film surface.

## I. Introduction

Metal-induced crystallization (MIC) has been investigated as a low-temperature process to deposit polycrystalline semiconductor materials on amorphous substrates [1-5]. For example, Al [1, 3], Au [4], and Ag [5] have been shown to reduce the crystallization temperature of amorphous germanium (*a*-Ge) to 150-250 °C, significantly lower than that for *a*-Ge solid phase crystallization [6]. Such a low-temperature crystal growth process enables the use of inexpensive substrates, e.g. large-area glass or flexible polymer sheets which are of interest for large-area electronics and solar photovoltaic technologies.

Polycrystalline germanium (*poly*-Ge) thin films have increasingly been studied for advanced large-area electronics such as thin-film transistors [7] and thin-film solar cells [8]. *Poly*-Ge thin-film transistors (TFTs) with Schottky source/drain contacts have been fabricated on glass, and good p-channel operation was demonstrated.[7] Ge has excellent lattice matching with GaAs and *poly*-Ge films can serve as epitaxial templates for two-dimensional growth of polycrystalline GaAs layers. Interesting device applications of such layered stacks may be found in polycrystalline GaAs single junction or GaAs/Ge multi-junction solar cell.[8] Furthermore, (111) single crystal Ge substrates have been used to grow vertical-aligned semiconductor nanowire arrays [9, 10]. Therefore, textured *poly*-Ge films with (111)-preferred orientations deposited on glass or polymer sheets can be used as a growth template for assembling aligned nanowire or microwire photovoltaic devices on large area substrates [11-13].

We recently demonstrated Al-induced layer exchange crystallization to form continuous *poly*-Ge thin films with micron-sized grains and (111)-preferred orientation at 200 °C. A sub-nm $GeO_x$ ($1<x<2$) interfacial layer, which is engineered to provide relatively sparse, nanoscale fast-diffusion paths for Ge, is critical for controlling nucleation of (111)-oriented Ge crystallites on the initially-overlying Al film. With a well-controlled interfacial layer, the Al/Ge bi-layer structure can serve as a model system to study layer exchange crystal growth kinetics in metal-induced crystallization. Understanding and controlling the kinetics of MIC at the nanoscale is important for achieving semiconductor thin films with tunable properties and practical success in manufacturing.

In this paper, Ge layer exchange crystallization controlled by a $GeO_x$ interfacial layer is observed directly by imaging with elemental contrast using secondary electron microscopy. We present a systematic study of Ge crystallization kinetics for various annealing temperatures. Based on independent analysis of nucleation and growth, we have used Johnson-Mehl-Avrami-Kologoromov (JMAK) phase transformation theory [14] to model *poly*-Ge nucleation and lateral growth in two dimensions. The Ge island growth kinetics switch from Ge atomic-attachment-limited growth at the amorphous Ge (*a*-Ge)/crystalline Ge (c-Ge) interface to Ge surface diffusion-limited growth as the growth front advances across the Al surface. The transition point is sensitive to the density of nucleated Ge islands and the annealing temperature. We demonstrate that bias-voltage stressing to induce dielectric breakdown of the $GeO_x$ layer enhances the nucleation density of Ge islands during layer exchange.

**II. Experiments**

Amorphous germanium (*a*-Ge) and crystalline Al (c-Al) films were deposited in an electron-beam evaporation system with no intentional substrate heating. First, 100 nm thick *a*-Ge thin films were thermally evaporated on $SiO_2$ substrates. Substrates were Si (100) wafers with a 100-nm-thick thermally oxidized $SiO_2$ film. A series of cleaning procedures were used to remove hydrocarbon and metal contaminants prior to *a*-Ge film deposition. These include a sequence of isopropyl alcohol (IPA) rinse for 5 minutes, 4:1 $H_2SO_4:H_2O_2$ dip cleaning at room temperature for 10 minutes, 5:1:1 $H_2O:H_2O_2:HCl$ dip cleaning at room temperature for 10 minutes, followed by drying in $N_2$ atmosphere. The sub-nm $GeO_x$ interfacial layer was prepared by flowing ozone gas over the as-deposited *a*-Ge film to oxidize the film surface without substrate heating. The $GeO_x$ thickness and oxygen stoichiometry x is determined by the oxidization time, using an $O_3$ partial pressure fixed at 0.24 Torr. Oxidized samples were immediately transferred to the electron-beam evaporation system and used as the substrate for Al layer deposition. The thickness of blanket Al films (for crystallization kinetics study) and patterned Al pads (for bias stressing experiments) was 50 nm. Samples with the same c-Al, *a*-Ge and $GeO_x$ layer characteristics were fabricated in batches.

Annealing was carried out on a heating stage at temperatures from 200 °C to 300 °C in ultra high vacuum. For each Al/*a*-Ge bi-layer sample, the annealing time ranged from 15 minutes to 3 hours. The resistive heater attains the annealing temperature within minutes, and anneals were terminated by removing the sample from vacuum and cooling them down within 1 minute in flowing $N_2$. Plan-view characterization methods were employed to study Ge crystallization kinetics: Auger elemental mapping was acquired with a PHI 700 scanning Auger system at 10keV (probe size: 15 nm); scanning electron microscopy (SEM) images were from an FEI XL 30 Sirion SEM, with an electron beam at 3keV and a conventional secondary electron detector.

We have also studied the effect of bias-voltage stressing of the oxide interfacial layer. In these experiments, a 5 nm Ti layer and a 25 nm Pt layer were deposited sequentially as the back electrode prior to 100 nm *a*-Ge deposition, and Al films were patterned into 200 μm diameter circular pads. A computer-programmed voltage source applied a constant bias between the Al pad and the underlying Pt/Ti metal layers. The voltage drop across the $GeO_x$ layer was defined by the applied bias. The leakage current across the oxide layer was measured by a current meter and recorded automatically.

**III. Results and Discussion**

Al-induced layer exchange crystallization of *a*-Ge was directly observed by imaging Ge islands on Al with the elemental contrast after thermal annealing. During thermal annealing, Ge atoms diffuse upwards, through the interfacial $GeO_x$ layer, to the Al surface, nucleate on the Al surface and overgrow laterally until impinging with one another, resulting in a distribution of Al and Ge elements on the sample surface prior to completion of *a*-Ge crystallization. Figure 1 shows SEM micrographs and Ge and Al Auger elemental maps. Figure 1(a)-(c) are from the same location of the sample after 90 minutes annealing at 250 ℃. The bright contrast regions in figure 1(a) are rich in Ge, corresponding to the green area in figure 1(b); the dark contrast regions in figure 1(a) are Al-rich, corresponding to the red area in figure 1(c). Figure 1(d)-(e) show near-complete Ge surface coverage after annealing for 3 h at 250 ℃. Given sufficiently long annealing time, crystalline Ge can substantially cover the original Al surface.

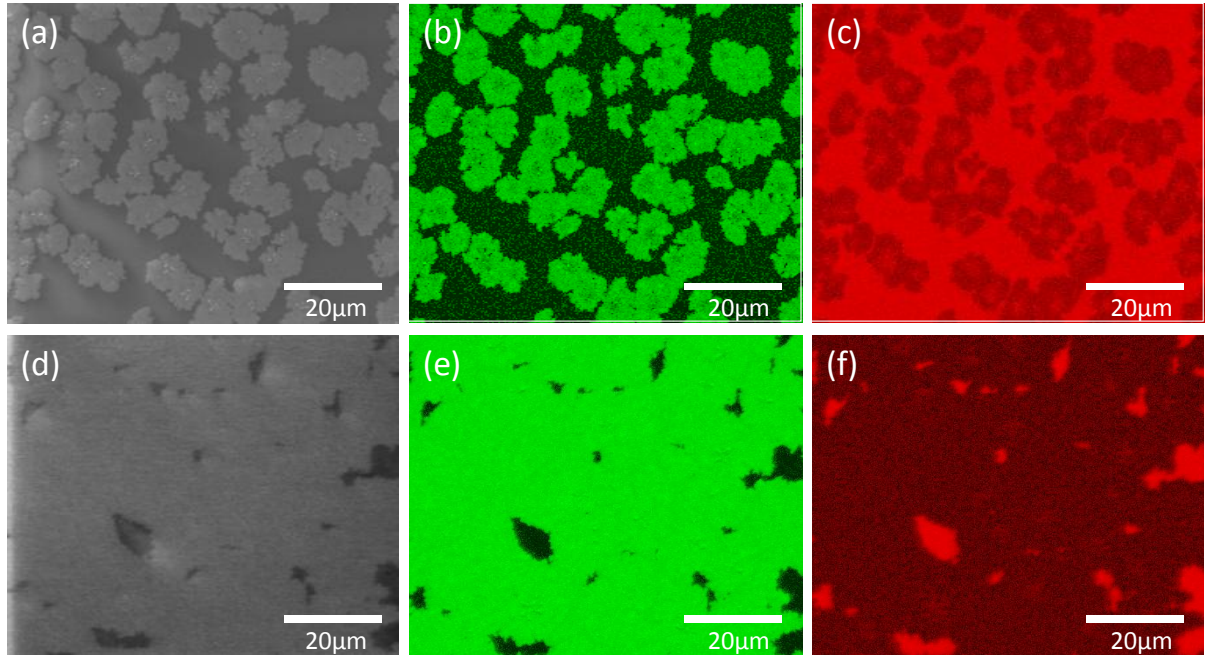

**Figure 1.** Surface composition analysis of the Al/GeO$_x$/Ge samples after 250 ℃ annealing, (a)-(c) for 90 min and (d)-(f) for 3 h. (a), (d) Plan-view SEM micrographs (recorded in the scanning Auger microscope) for the samples annealed for 90 min and 3h, respectively. (b), (c) Ge and Al elemental maps from the same location of the sample in (a). (e), (f) Ge and Al elemental maps from the same location of the sample in (d).

Scanning electron microscopy was used to characterize Ge surface coverage on Al for all the annealed samples in the following study. Imaging the Ge surface distribution at various stages of annealing allows us to independently measure the density and average area of Ge crystal islands as a function of annealing time. The sizes of Ge islands are correlated with the sizes of Ge grains, as shown in supplementary materials. The thickness of Ge islands remains constant as they grow radially, as indicated by the uniform thickness of as-crystallized Ge layers in the cross-section view [1]. Figure 2 shows that Ge surface coverage, or the area fraction of Ge on Al, increases after annealing at 250 ℃ for 15, 30, 45, 60, 90, 120, and 180 minutes. The SEM images, with one displayed per sample, were from two sample batches: one batch with a thin $GeO_x$ diffusion control layer prepared by 35 s $O_3$-oxidation; and the other with a thicker layer prepared by 45 s $O_3$-oxidation. Detailed analysis of the images indicates how Ge island nucleation and growth behave as $GeO_x$ layer characteristics are varied and as a function of annealing conditions at various stages of crystallization.

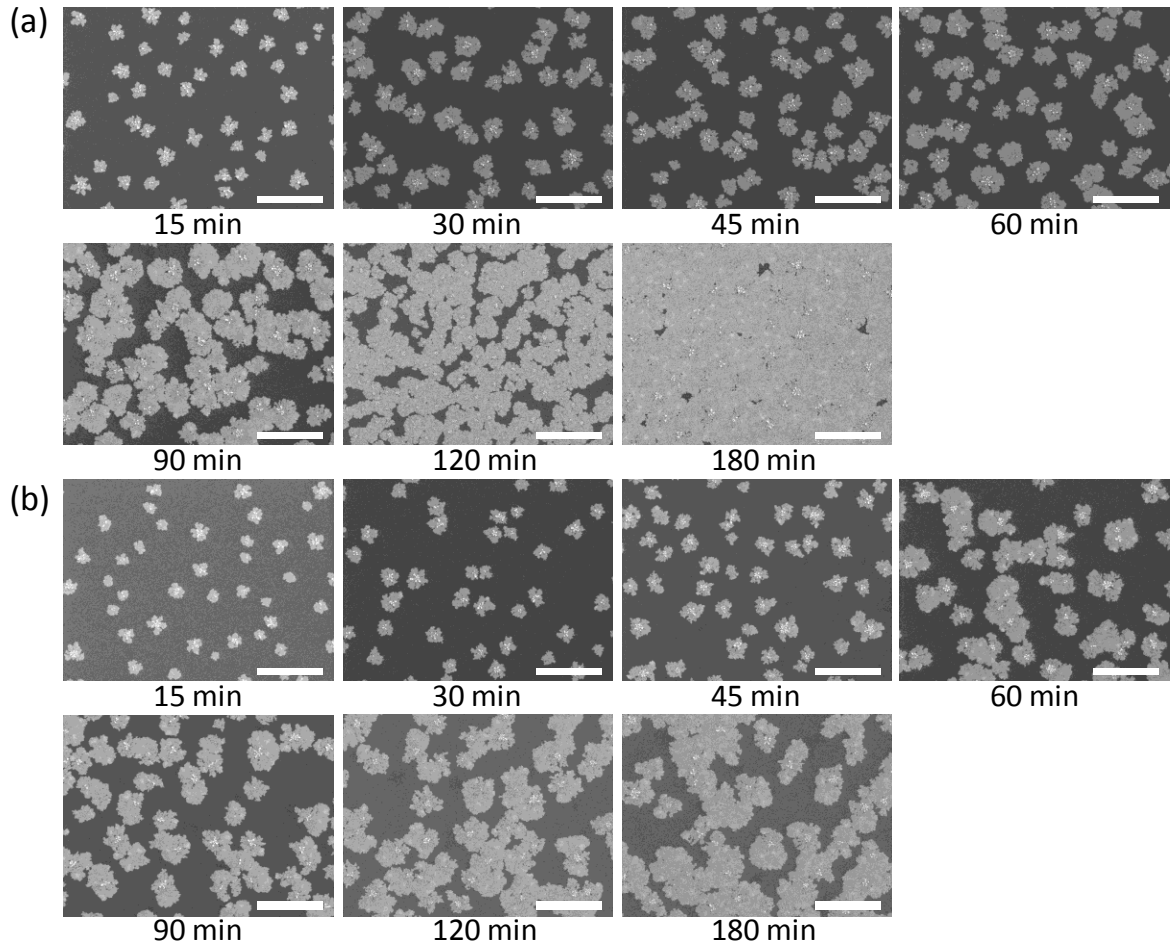

**Figure 2.** Two series of SEM micrographs show the surface coverage of crystalline Ge islands vs. annealing time at 250 ℃, for samples (a) with a thin $GeO_x$ diffusion control layer (35 s $O_3$-oxidation) and (b) with a thicker layer (45 s $O_3$-oxidation). Scale bar for all images is 20 μm.

After thermal annealing, the density of Ge crystal islands on Al was measured and then interpreted with the proposed kinetics model. The images, which were collected from random locations on each sample, were analyzed with the software package ImageJ (developed at National Institute of Health). The bright contrast regions were marked and counted as crystallized Ge islands, in keeping with the contrast differences observed in figure 1. Each data point is obtained by statistical analysis of the image data. Figure 3(a) shows the nucleation behaviour for different interfacial layers ($GeO_x$ by $O_3$-oxidation for 35 s vs. 45 s) with 250 °C layer exchange annealing. There is an upper limit for the density of Ge islands, which saturates at $120.6 \times 10^4$ /cm$^2$ for the thin $GeO_x$ layer and $92.1 \times 10^4$ /cm$^2$ for the thicker layer, after ~30 min and ~60 min, respectively. Assuming that there were a fixed number of randomly distributed sites where Ge islands could possibly nucleate, the saturated density of these sites is correlated with the areal density of fast-diffusion paths through the $GeO_x$ layer. Maximum nucleation density was achieved within 10's of minutes of annealing, which suggests a decaying nucleation rate with time. The growth of individual Ge islands is also studied for samples with both interfacial layer thicknesses, as shown in figure 3(b) and 3(c). Below, a theory of nucleation, growth and impingement is described, and then will guide the subsequent analysis.

We model such layer exchange crystallization as an isothermal solid phase transformation with nucleation of Ge islands and their radial growth in two dimensions [1]. An analytical description according to Johnson, Mehl, Avrami, and Kologoromov (JMAK) [14] is applicable to this case. In this model, the Ge island nucleation rate can be taken on one of several different time dependences in order to the observed evolution of nucleus density on the sample surface. The nucleation can be site saturation, continuous nucleation and Avrami nucleation, or a mix of above, which is determined by time dependence of the observed nuclei density. As measured in our experiments, the decaying nucleation rate with time indicates Avrami nucleation.

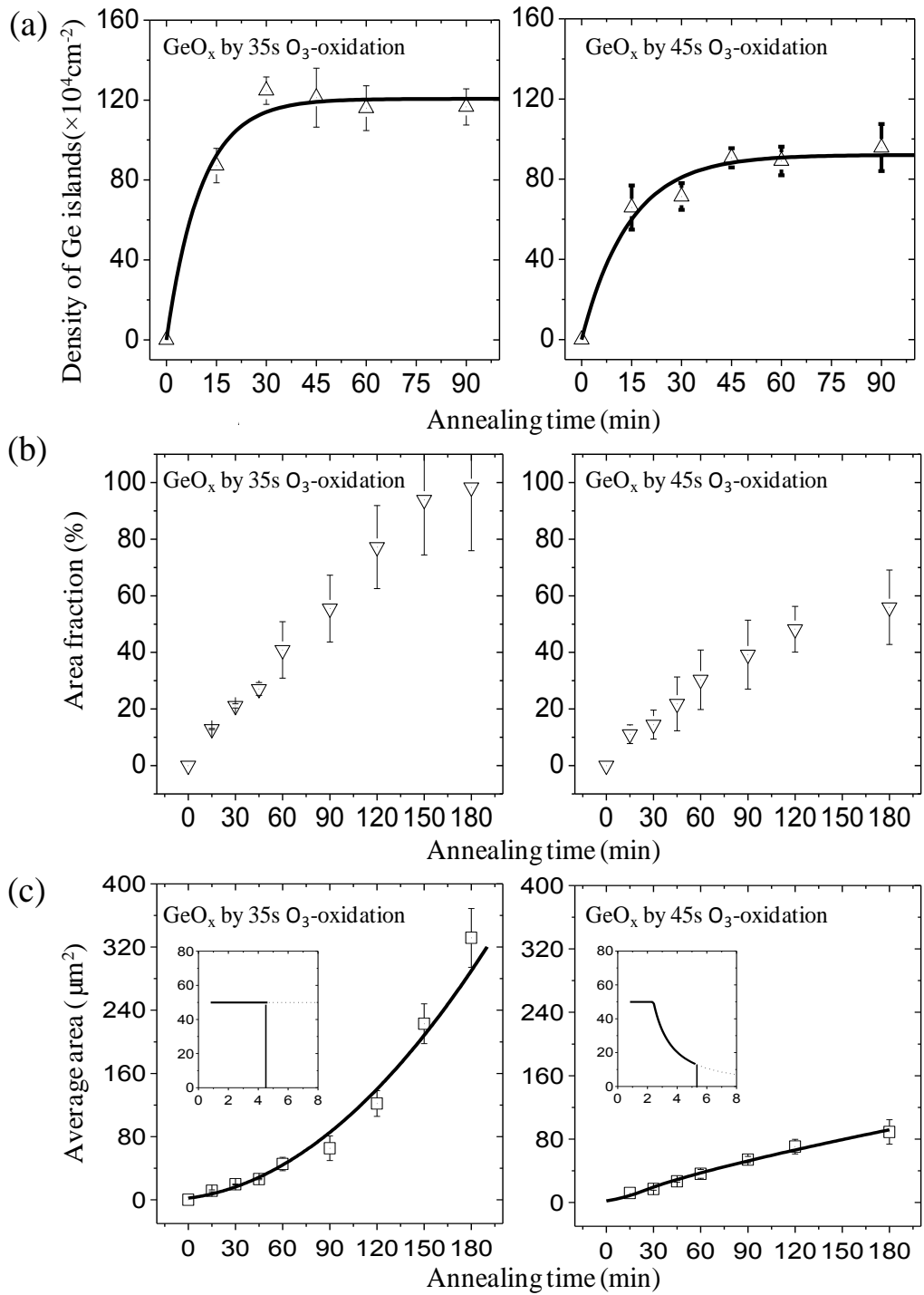

**Figure 3.** Nucleation and growth behaviours of Ge crystal islands at 250 °C annealing, for the samples shown in figure 2. (a) The density, (b) Ge surface coverage on Al and (c) average area of Ge islands as a function of annealing time for both samples. The insets in (c) plot calculated growth velocities (nm/min) as a function of Ge island radii (μm), with the dashed line indicating the occurrence of substantial impingement.

For growth of individual Ge islands, we consider two limiting cases: 1) Ge atomic migration across the *a*-Ge/c-Ge interface at the island growth front and 2) Ge surface diffusion-limited growth, characterized by Ge surface atomic fluence that decreases with the island radius and annealing time. When there is a transition between these two growth regimes, one expects Ge islands to initially grow radially at a constant rate $v_G$ up to a threshold radius $r_T$, with further growth being diffusion-limited. Our simple model assumes that Ge islands nucleate with a circular shape at a critical radius $r_e$ and that the initial growth rate $v_G$ is only temperature dependent with an activation energy of $Q_G$, which is related to the energy barrier for Ge atoms to migrate from the amorphous to the crystalline phase. The film thickness across a growing Ge island is approximately uniform. Under diffusion-limited conditions, the Ge island lateral growth velocity will become:

$$v = \frac{dr}{dt} = \frac{\mu_\alpha - \mu^*}{\mu^* - \mu_c} \frac{D}{r \ln(r/r_e)} = \frac{D^*}{r \ln(r/r_e)}, \quad (1)$$

where $D$ is the surface or interface diffusion coefficient for Ge atoms diffusion on the Al film surface or along the Al/Ge interface, and $\mu_c$, $\mu^*$, and $\mu_\alpha$ stand for the chemical potential of Ge atoms inside the island, at the growth front, and inside the amorphous Ge phase far from the growth front. Because the growth curves shown in figure 3(c) are mostly non-linear with time, we ignore the case of a constant Ge atomic fluence to the island growth front (e.g. growth that is limited by vertical transport of Ge atoms from the *a*-Ge layer to the Al surface). Detailed derivation of the diffusion-limited radial growth of a cylindrical shape island can be found in reference [14]. We cannot ignore the interference between neighboring Ge islands that compete for the sources of amorphous Ge, which might change the value of $\mu_\alpha$. However, the values of $\mu_\alpha$ for different GeO$_x$ layers may not be comparable only when the density of nucleated Ge islands varies by orders of magnitude. Define the *effective diffusion coefficient*,

$D^*$, as $D^* = D \cdot (\mu_\alpha - \mu^*)/(\mu^* - \mu_c)$. For a first-order approximation, $D^*$ is not expected to change much if the density of nucleated Ge islands is almost constant. Therefore, the growth of individual Ge islands can be described in the following compact form. Continuity of island growth velocity at $r = r_T$ is imposed to combine the two growth mechanisms, resulting in:

$$v = \frac{dr}{dt} = \begin{cases} v_0 \exp(-Q_G/kT), & r \leq r_T \\ \frac{\mu_\alpha - \mu^*}{\mu^* - \mu_c} \frac{D}{r \ln(r/r_e)}, & r > r_T \end{cases} = \begin{cases} v_G, & r \leq r_T \\ \frac{D^*}{r \ln(r/r_e)}, & r > r_T \end{cases}, \quad (2)$$

with the threshold radius $r_T$ satisfying,

$$\frac{D^*}{r_T \ln(r_T/r_e)} = v_G \quad (3)$$

$r$ is the radius of the circular growth front; $v_0$ is the temperature-independent pre-factor and $Q_G$ is the activation energy for the temperature-dependent growth rate $v_G$. The transition point to diffusion-limited growth will depend on comparison of the constant growth rate $v_G$ and effective surface diffusion coefficient $D^*$ at temperature $T$. Considering this impingement model [15], the area fraction of Ge on Al, which is the degree of layer exchange crystallization $f(t)$, is finally written as:

$$f(t) = 1 - \exp(-A^e/A) = 1 - \exp\left(-\int_0^t \dot{N}_N(\tau) \cdot Y(\tau,t) d\tau\right). \quad (4)$$

Two methods were used to track the time dependence of the average area of Ge islands, as a proxy measure for Ge island growth: (1) directly counting islands when there is no impingement, and (2) indirectly calculating from the surface area fraction covered and the areal density of Ge islands nucleated on the Al surface using Eq. (4), after impingement of Ge islands begins. The second method is a good estimate because the nucleation density quickly saturates ($\tau_N$ =10~15 min in Avrami nucleation) during annealing. The values measured by both methods are comparable when there is no impingement. Furthermore, the second method considers the contribution of Ge island growth to the increase of the *extended area* that is calculated from Ge surface coverage, as an important intermediate step for the kinetics analysis. Although the observed fractal growth of Ge islands in figure 2 and in reported Au/Ge crystallization [16] indicates local anisotropy of the growth velocity, our assumption of isotropic radial growth of Ge islands is in reasonably good agreement with the measured data. As shown in figure 3(c), the data points for the thin $GeO_x$ layer exhibit a parabolic increase of average island area with time, indicating a constant growth rate. Data for the thicker $GeO_x$ exhibit almost linear increase with time at later stages of annealing, indicating diffusion-limited Ge island growth.

Interestingly, one of two distinctive growth mechanisms appears to dominate after 60 min annealing, depending on the thickness of the $GeO_x$ interfacial layer. The maximum density for nucleated Ge islands is $120.6 \times 10^4$ vs. $92.1 \times 10^4 /cm^2$ for the $GeO_x$ layers prepared by 35 s vs. 45 s $O_3$-oxidation. As the nuclei density decreases, the average distance between neighbouring Ge islands increases. Without impingement considered, a Ge island should follow a transition from constant interface velocity (e.g. atomic-migration-limited) growth to Ge surface diffusion-limited growth, consistent with the results shown in Fig. 3(b) for the thicker $GeO_x$ layer. However, if the *poly*-Ge islands begin to impinge on one-another, neighbouring Ge islands impose geometric constraints on one-another's lateral growth. If the average island separation is small enough, e.g. for the samples with the thin $GeO_x$ layer, a substantial number of Ge islands coalesce prior to the transition to the diffusion-limited growth regime. In this case, the actual Ge surface area fraction on Al continues to increase (shown in figure 3(b)), as does the *extended area* fraction in a parabolic fashion with time. Figure 3(c) insets show the calculated growth velocities as a function of the Ge island radius. When the radius of Ge island growth front exceeds the average island separation, isotropic growth is no longer the case for most Ge islands, as indicated by the dashed line in the figure 3(c) insets. The threshold radius (~3 μm) is comparable with the average distances between neighbouring Ge islands for the 45 s $GeO_x$ sample. This comparability might explain why small change in saturation density results in change of growth mechanisms.

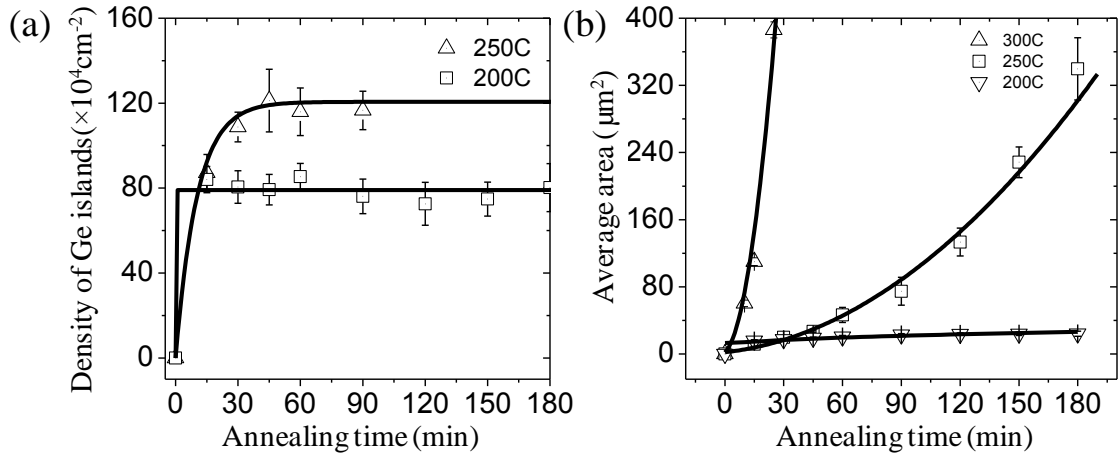

**Figure 4.** Temperature dependence of (a) nucleation and (b) growth kinetics of the Ge islands in the 35s $O_3$-oxidation samples.

In addition, the temperature dependence of Ge crystal nucleation and growth kinetics has been studied. We focused on the samples with 35s $O_3$-oxidation, and repeated the previous analysis for both 200 ℃ and 300 ℃ annealing. The measured data points are interpreted using the aforementioned model, with fitted parameters listed in table 1. The calculated curves for the density and average area of Ge islands are the least-squares fitting to the model and are also plotted in figure 3 and 4. Figure 4(a) shows the density of nucleated Ge islands vs. annealing time, with the maximum nucleation density equal to $79.1 \times 10^4$ /cm$^2$ for 200 ℃ annealing and with the Avrami time constant close to zero. Lower annealing temperature reduces the probability of a certain nucleation site being active, decreasing the maximum nucleation density. Figure 4(b) shows Ge island growth after 200 ℃, 250 ℃ and 300 ℃ annealing. The growth at 300 ℃ obeyed constant interface velocity (e.g. atomic-migration-limited) kinetics, with a constant radial island growth rate of $400 \pm 20$ nm/min. The activation energy for the interface-limited growth velocities is estimated to be $0.90 \pm 0.11$ eV. On the other hand, the growth at 200 ℃ exhibited a strongly diffusion-limited mechanism, with the transition to surface diffusion-limited growth occurring at an early stage of annealing.

**Table 1.** List of fitted parameters for the kinetics model of nanoscale nucleation and lateral growth of Ge islands in two dimensions.

| Sample and annealing condition | Maximum nucleation density, $N_0$ ($10^4 cm^{-2}$) | Nucleation saturation time, $\tau$ (minutes) | Growth Mechanism | Interface-limited growth velocity, $v_0$ (nm min$^{-1}$) | Diffusion-limited growth | |
|---|---|---|---|---|---|---|
| | | | | | Effective diffusion coefficient, $D^*$ (cm$^2$ s$^{-1}$) | Size of critical nuclei, $r_e$ (μm) |
| GeO$_x$ by O$_3$-oxidation for 35 s / 250°C | 117.7±4.5 | 11.5 | Interface-limited | 51±1 | Not applicable | |
| GeO$_x$ by O$_3$-oxidation for 45 s / 250°C | 92.1±4.2 | 14.3 | Interface-limited /diffusion-limited | 51±1 | (2.13±0.25) ×10$^{-11}$ [a] | 0.83±0.01 |
| GeO$_x$ by O$_3$-oxidation for 35 s / 200°C | 79.1±3.5 | $\tau \to 0$ | Mostly diffusion-limited | 7.9±0.8 | (4.28±0.17) ×10$^{-13}$ [a] | 2.0±0.1 |
| GeO$_x$ by O$_3$-oxidation for 35 s / 300°C | >150 | $\tau \to 0$ | Interface-limited | 400±20 | Not applicable | |

[a] The lower limits for the surface diffusion coefficients are estimated.

The disparity in growth mechanisms for 250 °C and 300 °C vs. 200 °C is due to the temperature dependence of the threshold radius. Both constant growth rate process (e.g. Ge atom attachment at the island growth front) and Ge surface diffusion on the Al layer are thermally activated processes. Analysis of the data in figure 4 indicates that when the annealing temperature decreases from 250 °C to 200 °C, the constant growth rate, $v_G$, reduces by a factor of ~1/6, while the effective surface diffusion coefficient, $D^*$, decreases by ~2 orders of magnitude. The diffusion-limited growth velocity decreases much more strongly with temperature reduction, so that the threshold radius shifts to a smaller value for lower annealing temperatures. With less nucleation density at 200 °C, the growth of Ge islands is strongly surface diffusion-limited.

Finally, we also studied the effects of local bias-voltage induced dielectric breakdown of the GeO$_x$ diffusion barrier layers. This provides a way to tune the density of nucleated Ge islands with externally applied electrical field. Al pads (200 μm in diameter) were patterned on the *a*-Ge film with an O$_3$-grown GeO$_x$ layer. Then, 1.0V DC bias was applied to individual Al pads for different durations, with the sample annealed immediately thereafter at 200 °C for 20 minutes in vacuum to induce layer exchange. Statistically, there are two types of biasing histories. As shown in Figure 5, the black curve represents one type in which current spikes occur periodically (e.g. at times <140 s) and the leakage current density gradually increases, an indication of accumulated damage in the GeO$_x$ layer akin to soft dielectric breakdown[17]. The red curve represents the other type of biasing history, where the current density remains low before the occurrence of any significant change in current density, and then the bias is removed. The cumulative charge flowing across the GeO$_x$ layers for both types of biasing experiments is on the order of $10^{-5}$ C and $10^{-10}$ C, respectively. The density of nucleated Ge islands varies significantly as a function of the biasing history of the Al pads, as indicated by the series of SEM micrographs in figure 5(b). DC bias stressing of the GeO$_x$ layer increases the areal density of defective sites in the barrier layer, and thus controls the density of Ge islands on the Al film surface.

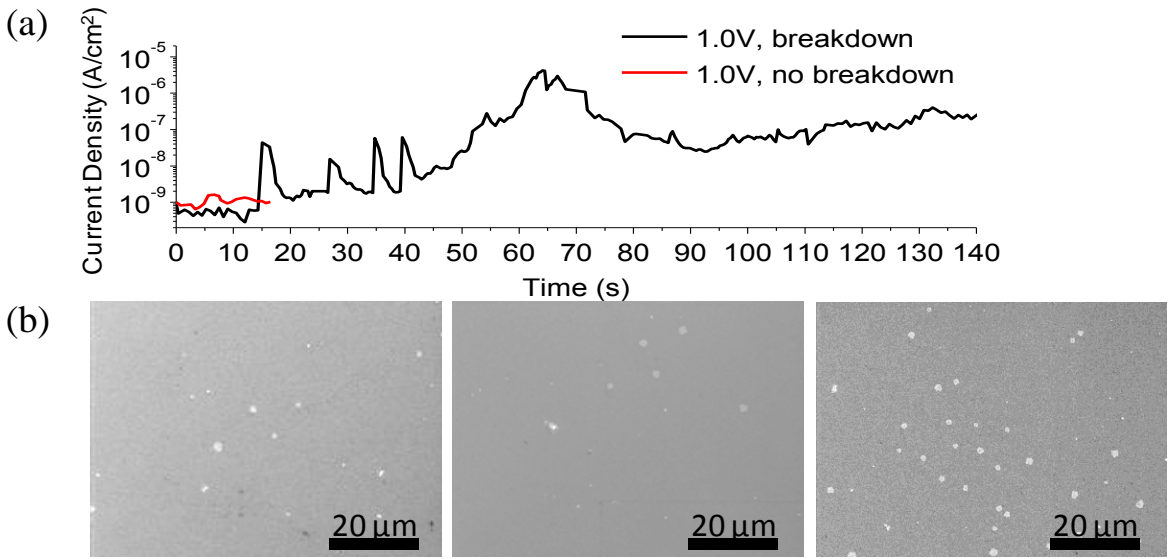

**Figure 5.** Bias-voltage induced dielectric breakdown of the GeO$_x$ diffusion control layer. (a) Current density vs. biasing time, with biasing histories showing the breakdown and non-breakdown at 1.0V. (b) A series of SEM micrographs for Al pads without a biasing history, with the red curve and with the black curve type of biasing histories shown in (a).

**IV. Conclusions**

We have directly observed nucleation and growth of Ge crystal islands during Al-induced layer exchange crystallization of *a*-Ge. Al/Ge bi-layers with a GeO$_x$ interfacial layer, of which the thickness and oxygen stoichiometry were controlled by *a*-Ge oxidation time, were examined in detail to study the kinetics of metal-induced layer exchange crystallization. SEM image analysis was used to quantify the area occupied by crystalline Ge islands on the Al surface at intermediate stages of annealing.

Based on the experimental observations, we have extended the JMAK isothermal phase transformation theory to model Ge island nucleation and growth kinetics in two dimensions. The density of nucleated Ge islands is found to follow Avrami nucleation during annealing. The maximum nucleation density and nucleation saturation time of Avrami nucleation are correlated with GeO$_x$ layer characteristics. The growth of individual Ge islands follows Ge atomic-attachment to diffusion-limited growth mechanism, with a threshold radius $r_T$ determining the transition. The threshold radius, to which the growth front advances with a constant velocity, is sensitive to the density of nucleated Ge islands and the annealing temperature. Finally, we show that local bias-voltage stressing can increase the areal density of defect sites in the GeO$_x$ barrier layer, and thus control the density of nucleated Ge islands on the film surface. Understanding the joint effects of nucleation, interface-limited growth and surface diffusion is important to achieving *poly*-Ge thin films with tunable surface coverage and grain sizes at low annealing temperatures.

**Acknowledgements**

The authors would like to acknowledge Dr. Charles Hitzman for the Auger Electron Spectroscopy analysis. Funding for this work was provided by Stanford GCEP exploratory project and a Stanford Graduate Fellowship.

**Supplementary materials**

1. Correlation of the sizes of Ge islands and Ge crystal grains

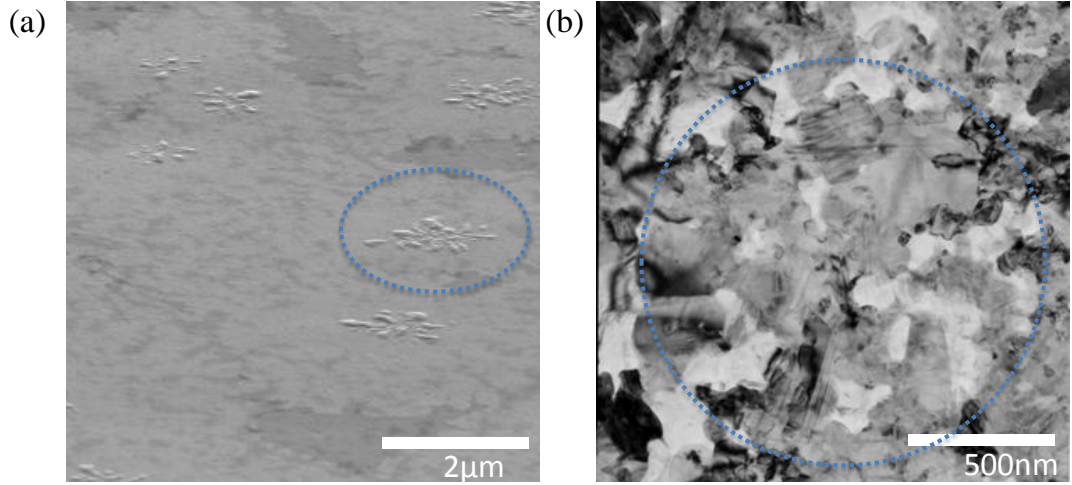

Transmission electron microscopy (TEM) was used to characterize Ge crystal grains in the *poly*-Ge films after layer exchange crystallization. SEM (a) and TEM (b) characterization of Ge crystal islands for the Al/GeO$_x$/Ge sample after 250 °C annealing for 2 h. The flower-shape features in both images indicate the correlation between the sizes of Ge islands and the sizes of Ge grains. TEM characterization was from an FEI Tecnai G2 F20 TEM at 200kV.

2. Time dependences of Ge island nucleation

The nucleation can be site saturation, continuous nucleation, Avrami nucleation, or a mix of above.

*Site saturation* is used where all nuclei are present at the beginning of transformation (*t*=0) already:

$$\dot{N}_N(T) = N^* \delta(t-0), \quad \quad \text{S(1)}$$

where $\delta(t-0)$ is the Dirac function. The Avrami nucleation involves the formation rate of supercritical nuclei at time *t*, assuming that the rate of forming new nuclei is proportional to the number of remaining nucleation sites.[18] If $\lambda$ is the nucleation rate of a given site, then:

$$N_N = N_0[1 - \exp(-\lambda t)] = N_0[1 - \exp(-t/\tau)] \quad \quad \text{S(2)}$$

$$I = \dot{N}_N = N_0 \lambda \exp(-\lambda t), \quad \quad \text{S(3)}$$

where $N_N$ is the nuclei density, $I$ is the nucleation rate, $N_0$ is the maximum nucleation density in Avrami nucleation, and $\tau_N$ is the characteristic nucleation saturation time where $\tau_N = 1/\lambda$. Avrami nucleation involves an exponentially decaying nucleation rate with time. It reduces to site saturation if $\lambda \to \infty$. On the other hand, if $\lambda \to 0$, eq. S(3) can be written as

$$I = \dot{N}_N = N_0 \lambda, \qquad \text{S(4)}$$

which is the *continuous nucleation* because the resulting nucleation rate $N_0 \lambda$ is time-independent. This is also called *slow nucleation*. The specific name *mixed nucleation* represents a combination of site saturation and continuous nucleation modes.

3. Impingement of growing Ge islands

During annealing, Ge islands nucleate and grow laterally in two dimensions, as a good description for individual growing Ge islands. At time t, the area of a Ge crystal island nucleated at time $\tau$ is given by,

$$Y(\tau,t) = \pi \left( \int_\tau^t v dt' \right)^2, \qquad \text{S (5)}$$

for two dimensional lateral growth assuming the growth velocity is isotropic and the island thickness is constant. Eq. S(5) assumes that each Ge crystal grows continuously, in absence of other growing crystals. In this hypothesis, the total area of all Ge crystals at time *t*, called the *extended area*, is given by

$$A^e(t) = \int_0^t A \cdot \dot{N}_N(\tau) \cdot Y(\tau,t) d\tau, \qquad \text{S(6)}$$

where A is the sample surface area. In fact, we should account for the overlap of Ge crystals and disregard nucleation in already transformed regions. Considering this impingement model [15], the area fraction of Ge on Al, which is the degree of layer exchange crystallization $f(t)$, is finally written as:

$$f(t) = 1 - \exp(-A^e / A) = 1 - \exp\left(-\int_0^t \dot{N}_N(\tau) \cdot Y(\tau,t) d\tau\right). \qquad (4)$$